\documentclass[amssymb,prb,showpacs,twocolumn]{revtex4}
\input{psfig.sty}
\usepackage{epsfig}
\usepackage{dcolumn}
\usepackage{amsmath}
\usepackage{epsfig}
\begin{document}

\title{Three-terminal transport through a quantum dot in the
Kondo regime: Conductance, dephasing and current-current correlations} 
\author{David S\'anchez}
\author{Rosa L\'opez}
\affiliation{D\'epartement de Physique Th\'eorique,
Universit\'e de Gen\`eve, CH-1211 Gen\`eve 4, Switzerland}
\date{\today}

\begin{abstract}
We investigate the nonequilibrium transport properties of a three-terminal
quantum dot in the strongly interacting limit. At low temperatures, a Kondo
resonance arises from the antiferromagnetic coupling between the localized 
electron in the quantum dot and the conduction electrons
in source and drain leads. It is known that the 
local density of states is accessible through the differential conductance
measured at the (weakly coupled) third lead. Here, we consider the multiterminal 
current-current correlations (shot noise and cross correlations
measured at two different terminals). We discuss the dependence 
of the current correlations on a number
of external parameters: bias voltage, magnetic field
and magnetization of the leads.
When the Kondo resonance is split by fixing the voltage bias
between two leads, the shot noise shows a nontrivial
dependence on the voltage applied to the third lead.
We show that the cross correlations of the current are more sensitive
than the conductance to the appearance of an external
magnetic field. When the leads are ferromagnetic and
their magnetizations point along opposite directions,
we find a reduction of the cross correlations. Moreover, we report on the 
effect of dephasing in the Kondo state for a two-terminal geometry 
when the third lead plays the role of a fictitious voltage probe. 
\end{abstract}

\maketitle

\section{Introduction}
The Kondo effect represents a distinguished example of strong
many-body correlations in condensed matter physics.\cite{hew93}
Over the last fifteen years, much effort has been made in understanding
the implications of the Kondo effect on the scattering
properties of phase-coherent conductors. Indeed, the electric
transport through a quantum dot connected to two terminals
becomes highly correlated when
the temperature $T$ is lowered below a characteristic
energy scale given by $k_B T_K$.\cite{ng88}
At equilibrium, the Kondo temperature $T_K$ depends
on the parameters of the system, i.e., the coupling of the dot
to the external leads due to tunneling, the dot onsite repulsion
(charging energy)
and the position of the resonant level relative to the Fermi
energy $E_F$. All of them can be tuned in a controlled way.\cite{gol98} 

In a quantum dot with a sufficiently large charging energy ($U\gg k_B T$)
and a single energy level well below $E_F$, the dynamics of the quasilocalized
electron becomes almost frozen. Therefore, a quantum dot can be viewed as an
artificial realization (at the nanoscale) of a magnetic impurity with spin $S=1/2$.
At very low temperatures ($T<T_K$), charge fluctuations in the dot
are suppressed and there arises an effective antiferromagnetic
interaction between the electrons of the reservoir and the
$S=1/2$ localized moment. Remarkably, the measured conductance 
reaches the maximum value for a quantum channel ($2e^2/h$) and 
the dot appears to be perfectly transparent when a small voltage $eV_{\rm sd}$
is applied between the source and the drain contacts.
Nevertheless, the coherent correlated motion of the delocalized
electrons forming the Kondo cloud can be disturbed when either
the bias voltage or the temperature are of the order of $T_K$. 
In such a case, the many-body wave function 
of the Kondo state is expected to suffer from \emph{dephasing},
leading to a decrease in the conductance. This issue has recently attracted
a lot of attention\cite{kam00,col01,ros03}.
In this work, we mimic in a phenomenological way
the effect of dephasing on the transport properties of a two-terminal
quantum dot in the Kondo regime 
by introducing a \emph{fictitious} voltage probe. 

Now, in the absence of dephasing, the building block of the Kondo resonance
is a narrow peak (of width $\sim k_B T_K$) around $E_F$
in the local density of states (LDOS) of the dot. 
However, full quantum-dot spectroscopy of the LDOS \emph{cannot} be accomplished
with a two-terminal transport setup. In particular, one cannot gain
experimental access to the predicted large voltage induced splitting
of the LDOS when $eV_{\rm sd}>k_B T_K$.\cite{mei93,kon96,ros01,fuj03}
A way to circumvent this problem is by
attaching a third lead, as demonstrated independently
by Sun and Guo\cite{sun01} and Lebanon and Schiller.\cite{leb01}
In subsequent laboratory work, De Franceschi {\it et al.}\cite{fra02} observed
a split Kondo resonance by employing a slightly modified
technique---one of the leads was replaced by a narrow wire driven out
of equilibrium where left and right moving carriers have
different electrochemical potentials.

Motivated in part by the works cited in the preceding paragraph,
we are concerned in this paper as well with the nonequilibrium Kondo physics
and the fluctuations of the current through a quantum dot attached to 
three leads. As is well known, the investigation of the current-current
correlations in mesoscopic conductors has been a fruitful area
of research.\cite{bla00} Nevertheless, there are still very scarce
applications to strongly correlated systems as the shot noise
is a purely {\em nonequilibrium} property,
and thus more difficult to treat.
Hershfield\cite{her92} calculates the zero-frequency
shot noise using perturbation theory in the charging
energy (valid when the Kondo correlations are not large; e.g., at $T>T_K$). 
Yamaguchi and Kawamura\cite{yam94} choose the tunneling part of the Hamiltonian
as the perturbing parameter. Ding and Ng\cite{din97}
study the frequency dependence
of the noise by means of the equation-of-motion method (also reliable
for  $T>T_K$). Meir and Golub\cite{mei02}
perform an exhaustive study of the
influence of bias voltage in the shot noise of a quantum dot
in the Kondo regime. Dong and Lei\cite{don02} discuss the effect
on the shot noise of
both external magnetic fields and particle-hole symmetry breaking.
Avishai {\it et al.}\cite{avi03} calculate the Fano factor when
the leads are $s$-wave superconductors whereas the case of
$p$-wave superconductivity is treated by Aono {\it et al}.\cite{aon03}
The authors\cite{lop03a} examine the behavior of the Fano factor
at zero temperature when the formation of the Kondo resonance
competes with the presence of ferromagnetic leads and spin-flip processes.
L\'opez {\it et al.}\cite{lop03b}
make use of the two-impurity Anderson Hamiltonian
to address the shot noise in double quantum dot systems.
To the best of our knowledge, a study of the current fluctuations
in a multiprobe Kondo impurity is still missing.
This is the gap we want to fill here.

In mesoscopic conductors, B\"uttiker\cite{but92}
shows that the sign of the current cross correlations depends
on the statistics of the carriers. It is positive (negative) for
bosons (fermions) due to statistical bunching (antibunching).
This statement is based on a series of assumptions
(e.g., zero-impedance external circuits, spin independent transport,
normal thermal leads).
Positive correlations can be found if these conditions
are not met (see Ref.~\onlinecite{but03}
for references on this subject).
Here, we just mention a few studies based on structures
involving a quantum dot.
Bagret and Nazarov\cite{bag03} consider a Coulomb-blockaded
quantum dot attached to
paramagnetic leads whereas the ferromagnetic case
and the spin-blockade case are treated by
Cottet {\it et al}.\cite{cot03} B\"orlin {\it et al.}\cite{bor02} and
Samuelsson and B\"uttiker\cite{sam02} examine
the cross correlations of a chaotic dot in the presence
of a superconducting lead. In the spin dependent case,
S\'anchez {\it et al.} \cite{san03} find that the sign of the cross
correlations is affected by Andreev cross reflections.
In the context of quantum computation,
measuring current cross correlations
have been shown\cite{bur00}
to yield a indirect identification of the existence
of streams of entangled particles.
Therefore, the cross correlations are a valuable tool in characterizing
the electron transport in phase-coherent conductors.

In this work, we consider electron transport through a strongly
interacting quantum dot attached to three leads.
Section~\ref{model} explains the theoretical framework
(slave-boson mean-field theory) we use
to compute the conductance and the current-current correlations.
We show that the expressions for the cross correlations 
may be inferred from scattering theory applied to a Breit-Wigner
resonance with renormalized parameters.
Section~\ref{results} is devoted to the results.
First, we assume that the third lead is a fictitious voltage
probe and investigate the effect of dephasing with increasing
coupling to the probe. Then, we consider that lead as a real
probe and relate the differential conductance measured at one electrode
with the LDOS of the artificial Kondo impurity.
We show next that the sign of the cross correlations of the current
is negative, as expected from the fermionic character of
the Kondo correlations at very low temperature.
Moreover, we discuss the effect of bias voltage, external magnetic fields,
and spin-polarized tunneling in the cross correlations.
We finish this section with an investigation
of the effect of spin polarized transport in the shot noise.
Finally, Sec.~\ref{conclusion} contains our conclusions.

\section{Model}
\label{model}
We model the electric transport through the quantum dot
using the Anderson Hamiltonian in the limit of large onsite
Coulomb interaction $U\to\infty$. This way we neglect double
occupancy in the dot and the Hamiltonian is written in terms
of the slave-boson language:\cite{col84}
\begin{eqnarray}
&&\mathcal{H}=
\sum_{k\alpha\sigma}\varepsilon_{k\alpha\sigma}
c_{k\alpha\sigma}^\dagger c_{k\alpha\sigma}
+\sum_{\sigma}\varepsilon_{0\sigma}f_{\sigma}^\dagger f_{\sigma} \nonumber\\
&&+\sum_{k\alpha\sigma} (
V_{k\alpha} c_{k\alpha\sigma}^\dagger b^\dagger f_{\sigma}+
{\rm H.c.} ) \nonumber\\
&&+\lambda ( b^\dagger b +\sum_{\sigma} f_{\sigma}^\dagger f_{\sigma} -1) \,,
\label{eq-hamsb}
\end{eqnarray}
where $c_{k\alpha\sigma}^\dagger$ ($c_{k\alpha\sigma}$)
is the creation (annihilation)
operator describing an electronic state $k$ with spin
$\sigma=\{\uparrow,\downarrow\}$ and energy dispersion
$\varepsilon_{k\alpha\sigma}$ in the lead $\alpha=\{1,2,3\}$,
$\varepsilon_{0\sigma}$ is the (spin-dependent) energy level
in the dot and $V_{k\alpha}$ is the coupling matrix element.
The original dot second-quantization operators
have been replaced in Eq.~(\ref{eq-hamsb}) by a combination
of the pseudofermion operator $f_{\sigma}$ and the boson
field $b$. Hopping off the dot is described by the process
$c_{k\alpha\sigma}^\dagger b^\dagger f_{\sigma}$: whenever
an electron is annihilated by $f_{\sigma}$, an empty state in the dot
is created by $b^\dagger$ and then $c_{k\alpha\sigma}^\dagger$
generates an electron with spin $\sigma$ in the conduction band
of contact $\alpha$. The boson operator $b$ ($b^\dagger$)
may be regarded as a projection operator onto the vacuum (empty)
state of the quantum dot.
To make sure that a state with double occupancy
is never created, a constraint with Lagrange multiplier $\lambda$
is added to the Hamiltonian.

The current operator $\hat{I}_\alpha$ that yields the electronic
flow from lead $\alpha$ is given by
\begin{equation}
\hat{I}_\alpha=\frac{i e}{\hbar} [ \hat{N}_\alpha,\mathcal{H} ] \,,
\label{cur1}
\end{equation}
where
$\hat{N}_\alpha=\sum_{k \sigma} c_{k\alpha\sigma}^\dagger c_{k\alpha\sigma}$.
The general form of the power spectrum of the current
fluctuations reads\cite{sym}
\begin{multline}
S_{\alpha\beta}(\omega)=2 \int d\tau \, e^{i\omega\tau}
\langle \{\delta\hat{I}_\alpha(\tau),
\delta\hat{I}_\beta(0) \} \rangle
\\
= 2 \int d\tau \, e^{i\omega\tau} \left[
\langle \{\hat{I}_\alpha(\tau),
\hat{I}_\beta(0) \} \rangle -
\langle \hat{I}_\alpha \rangle\langle \hat{I}_\beta \rangle
\right] \,,
\label{noise1}
\end{multline}
$\delta\hat{I}_\alpha=\hat{I}_\alpha-{I}_\alpha$
describing the fluctuations of the current away from its average value
$I_\alpha=\langle \hat{I}_\alpha \rangle$.
We are interested in the zero-frequency limit of
$S_{\alpha \beta}(\omega)$. Since the energy scale $k_B T_K$
in typical experiments is of the order of 100~mK, the frequencies
should be $\omega\lesssim 2.4$~GHz.
Moreover, we shall work at $T=0$ (see below) so that
the current will fluctuate due to quantum fluctuations only
(we disregard thermal fluctuations).
\begin{figure}
\centerline{
\epsfig{file=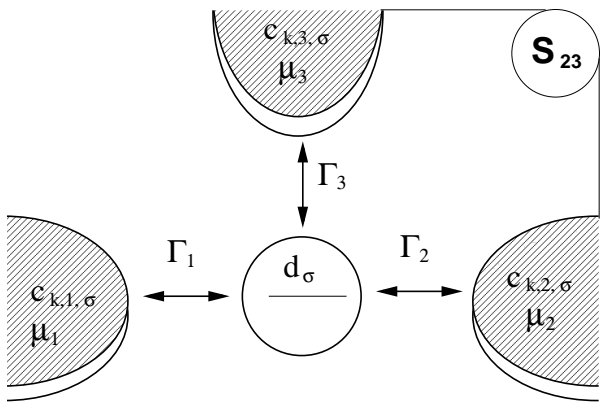,angle=0,width=0.45\textwidth,clip}
}
\caption{The system under consideration. The central island is a resonant
level coupled to three leads. The level may be shifted through a capacitative
coupling to a gate. In the limit of a vanishingly small capacitance,
double occupancy in the dot is forbidden and Kondo effect can arise.
The current--current cross correlations are measured between
leads 2 and 3.
}
\label{fig0}
\end{figure}

\subsection{Mean-field approximation}
The mean-field solution of the Hamiltonian~(\ref{eq-hamsb})
consists of considering the effect of the boson in an averaged way,
replacing the true operator $b(t)$ by its expectation value
$\langle b(t)\rangle$. Within this approximation the Hamiltonian
describes noninteracting quasiparticles with renormalized couplings:
$V_{k\alpha}\sqrt{|b|} \rightarrow \tilde{V}_{k\alpha} $.
The theory is then suitable for studying the Fermi-liquid fixed point
of the Kondo problem (i.e., at $T\ll T_K$) in which the averaged occupation
in the dot is always $1$. The dominant fluctuations in the
system are those associated to spin.

The stationary state of the boson field is determined from the
$t\to\infty$ limit of its equation of motion using the Keldysh
technique for systems out of equilibrium:\cite{lan76,hau98}
\begin{equation}
\sum_{k \alpha\sigma} \tilde{V}_{k\alpha}
G^{<}_{f_\sigma,k\alpha\sigma}(t,t)=-i \lambda |b|^2 \,,
\label{eomb}
\end{equation}
where $G^{<}_{f_\sigma,k\alpha\sigma}(t,t)=
i \langle c_{k\alpha\sigma}^\dagger (t) f_\sigma (t) \rangle$
is the lead-dot lesser Green function.
Next, we take into account the constraint:
\begin{equation}
\sum_{\sigma}
G^{<}_{f_\sigma,f_\sigma}(t,t)=i (1-|b|^2) \,,
\label{constraint}
\end{equation}
$G^{<}_{f_\sigma,f_\sigma}(t,t)=
i \langle f_{\sigma}^\dagger (t) f_\sigma (t) \rangle$
being the dot lesser Green function.
It yields the nonequilibrium distribution function in the dot.

In evaluating the above Green functions we need the coupling
strength given by $\Gamma_{\alpha\sigma}(\epsilon)=
\pi\sum_{k} |V_{k\alpha}|^2 \delta
(\epsilon-\varepsilon_{k\alpha\sigma})$. In the wide band limit,
one neglects the energy dependence of $\Gamma$ and 
the hybridization width is taken as
$\Gamma_{\alpha\sigma}=\Gamma_{\alpha\sigma}(E_F)$
for $-D\leq\varepsilon\leq D$
($D$ is the high-energy cutoff).
We notice that in the presence of Kondo correlations the 
lifetime broadening becomes renormalized 
$\Gamma_{\alpha\sigma}\rightarrow\tilde{\Gamma}_{\alpha\sigma}=
\pi\sum_{k} |\tilde{V}_{k\alpha}|^2 \delta
(\epsilon-\varepsilon_{k\alpha\sigma})$
and the bare level $\varepsilon_{0\sigma}$ is shifted to
$\tilde{\varepsilon}_{0\sigma}=\varepsilon_{0\sigma}+\lambda$.
We can now give the full expression of the Fourier-transformed
lesser Green function:
\begin{equation}
G^{<}_{f_\sigma,f_\sigma}(\epsilon)=
2i \frac{\sum_{\alpha}\tilde{\Gamma}_{\alpha\sigma} f_\alpha(\epsilon)}
{(\epsilon-\tilde{\varepsilon}_{0\sigma})^2+\tilde{\Gamma}_\sigma^2}\,,
\end{equation}
where $\tilde{\Gamma}_\sigma=\sum_\alpha \tilde{\Gamma}_{\alpha\sigma}$
is the total hybridization width per spin and
$f_\alpha(\epsilon)=\theta(\mu_\alpha-\epsilon)$
is the Fermi function at zero temperature
of lead $\alpha$ with electrochemical potential
$\mu_\alpha=E_F+eV_\alpha$.
On the other hand, 
$G_{f_\sigma,k\alpha\sigma}^{<} (\omega)$
can be cast in terms of
$G_{f_\sigma,f_\sigma}^{<} (\omega)$
with the help of the equation of motion of the operators
and then applying the analytical continuation rules
in a complex time contour.\cite{lan76}
Therefore, we obtain a closed system of two nonlinear equations
[Eqs.~(\ref{eomb}) and~(\ref{constraint})]
with unknowns $|b|^2$ and $\lambda$ to be found self-consistently.

From the precedent arguments and Eq.~(\ref{cur1})
we can easily establish an expression
for the expectation value of the electric current:
\begin{equation}
I_\alpha= \frac{e}{h} \sum_{\beta \sigma}
\int d\epsilon \, \tilde{T}_{\alpha \beta}^\sigma (\epsilon)
[f_\alpha(\epsilon)-f_\beta(\epsilon)] \,,
\label{cur2}
\end{equation}
which has exactly the same transparent form as
the Landauer-B\"uttiker formula\cite{but86} in the two channel
(one per spin) case applied to a double-barrier resonant-tunneling
system:
\begin{equation}
\tilde{T}_{\alpha \beta}^\sigma (\epsilon)= 
\frac{4\tilde{\Gamma}_{\alpha\sigma}\tilde{\Gamma}_{\beta\sigma}}
{(\epsilon-\tilde{\varepsilon}_{0\sigma})^2+\tilde{\Gamma}_\sigma^2}\,,
\label{cur3}
\end{equation}
which has a simple Breit-Wigner lineshape. For the same reason,
the quasiparticle density of states is a Lorentzian function
centered around the Fermi level (the Abrikosov-Suhl resonance).
This result is expected since we are dealing with
a Fermi liquid but we stress that the physics
it contains should \emph{not}
be confused with a noninteracting quantum dot since:

(i) $\tilde{T}$ depends implicitly on $|b|^2$ and $\lambda$, and it must
then be self-consistently calculated for each set of parameters:
contact voltages $\{V_\gamma\}$,
magnetic field $\Delta_Z=\varepsilon_{0\uparrow}-\varepsilon_{0\downarrow}$,
gate voltage $\varepsilon_{0}(V_g)$, and lead magnetization.

(ii) $\tilde{T}$ is renormalized by Kondo correlations (as the bare
$\Gamma$ and $\varepsilon_0$ are),

(iii) $\tilde{T}$ has a nontrivial dependence on the bias voltage.

All these features give rise to a number of effects that are not seen
in a noninteracting resonant-tunneling diode. There are many instances:
regions of negative differential conductance in the current--voltage
characteristics of a double quantum dot,\cite{agu00}
a crossover from Kondo physics to an antiferromagnetic
singlet in the two-impurity problem,\cite{lop03b}
an anomalous sign of the zero-bias magnetoresistance,\cite{lop03a} etc.
Below, we shall discuss another example without counterpart
in a noninteracting Breit-Wigner resonance: When the Kondo peak
splits due to a large bias voltage.

\subsection{Current-current correlations}
We consider now the current fluctuations given by Eq.~(\ref{noise1})
at zero frequency $S_{\alpha \beta}(0)$.
To simplify the notation we introduce
$G_0(\omega)=G_{f_{\sigma},f_{\sigma}}(\omega)$ as the dot Green function.
After lengthy algebra, we have
\begin{multline}
S_{\alpha \beta}(0) = \frac{4e^2}{h} \int d\epsilon \,
\tilde{\Gamma}_\alpha \tilde{\Gamma}_\beta [
G_0^{<} G_0^{>} -G_0^{a} G_0^{>} f_\alpha \\
+ G_0^{<} G_0^{a} (1-f_\beta)
-G_0^{<} G_0^{r} (1-f_\alpha) +G_0^{r} G_0^{>} f_\beta
-G_0^{a} G_0^{a} f_\alpha (1-f_\beta) \\
-G_0^{r} G_0^{r} f_\beta (1-f_\alpha)
-i\frac{\delta_{\alpha \beta}}{\pi\tilde{\Gamma}_\alpha}
(G_0^{<} (1-f_\beta) -  G_0^{>}f_\alpha ) ] \,.
\label{noise0}
\end{multline}

This formula (or variations of it) has been already employed in
the literature. Wei {\it et al.}\cite{wei99} prove it using the
Fisher-Lee-Baranger-Stone relation\cite{fis81}
to write the scattering matrix elements in terms of the retarded Green
function of the dot, $G_0^{r}$.
Dong and Lei\cite{don02} and L\'opez {\it et al.}\cite{lop03a,lop03b}
find it in Kondo problems within a slave-boson mean-field framework.
Actually, in Ref.~\onlinecite{lop03b} it is shown that
the shot noise in a two-terminal geometry reads 
$S\sim \tilde{T}(1-\tilde{T})$, i.e., the well known result for the
partition noise but with {\em renormalized} transmissions. 
Souza {\it et al.}\cite{sou02}
calculate the noise of an ultrasmall magnetic tunnel
junction by means of Eq.~(\ref{noise0}) within a Hartree-Fock framework.
In general, we can say that
Eq.~(\ref{noise0}) is consistent within mean-field theories.
However, some caution is needed if one wishes to go beyond a mean-field level.
In deriving Eq.~(\ref{noise0}), one needs to apply Wick theorem, which
is valid only for \emph{noninteracting} (quasi)-particles.
More specifically, one finds terms that read:
\begin{multline}
\langle c_{k\alpha\sigma}^\dagger (t) f_\sigma (t)
c_{k\beta\sigma'}^\dagger (0)  f_{\sigma'} (0) \rangle =\\
\langle c_{k\alpha\sigma}^\dagger (t) f_\sigma (t) \rangle
\langle c_{k\beta\sigma'}^\dagger (0) f_{\sigma'} (0) \rangle\\
+\langle c_{k\alpha\sigma}^\dagger (t) f_{\sigma'} (0) \rangle
\langle f_\sigma (t) c_{k\beta\sigma'}^\dagger (0)  \rangle \,.
\end{multline}
The first term in the left-hand side corresponds to disconnected diagrams
that cancel out the term
$\langle \hat{I}_\alpha \rangle\langle \hat{I}_\beta \rangle$
of Eq.~(\ref{noise1}) whereas the second term contributes to
Eq.~(\ref{noise0}).
Therefore, the particular
Hamiltonian has to be cast first in a quadratic form.
Zhu and Balatsky\cite{zhu03}
incorrectly state that Eq.~(\ref{noise0}) takes into account the many-body
effects. Also, it is not clear how this formula is inferred within
the equation-of-motion method employed by L\"u and Liu.\cite{lu02}

In our case, the mean-field approximation is known to be
the leading term in a $1/N$ expansion,\cite{new87}
where $N=2$ is the spin degeneracy.
Therefore, we neglect the fluctuations of both the boson field ($\delta b=0$)
and the renormalization of the resonant level ($\delta \lambda=0$),\cite{don02,col84}
which could be calculated in the next order.
This is valid as long as we restrict ourselves to the Fermi-liquid fixed
point of the Kondo problem. We are not aware
of real $1/N$ correction calculations of shot noise. 
Although Meir and Golub\cite{mei02} perform a noncrossing approximation (NCA),
they just substitute the NCA propagators into Eq.~(\ref{noise0}),
with the limitations exposed above.

The current-current correlations can be deduced either using Eq.~(\ref{noise0})
or using the scattering approach for the multiterminal case
(see Ref.~\onlinecite{but92}). The latter formalism amounts to replacing the bare
parameters by the renormalized ones~\cite{lop03b}. We consider the
illustrative case of having different electrochemical
potentials in two leads, $\mu_\alpha\neq\mu_\beta$ (e.g., $\alpha=2$ 
and $\beta=3$) at zero temperature. We find
\begin{equation}
S_{23}(0) = -\frac{2e^2}{h} \sum_{\gamma,\delta} \int d\epsilon \,
{\rm Tr} (s_{2 \gamma}^\dagger s_{2 \delta} s_{3 \delta}^\dagger s_{3 \gamma})
(f_\gamma -f_a) (f_\delta -f_b)\,,
\label{ccc2}
\end{equation}
where $s_{\alpha\beta}$ is the renormalized scattering amplitude of a
Breit-Wigner resonance:
\begin{equation}
s_{\alpha\beta}^\sigma(\epsilon)=\delta_{\alpha\beta}-
\frac{2i\sqrt{\tilde{\Gamma}_{\alpha\sigma}\tilde{\Gamma}_{\beta\sigma}}}
{\epsilon-\tilde{\varepsilon}_{0\sigma}+i\tilde{\Gamma}_{\sigma}}\,.
\label{bw}
\end{equation}
In Eq.~(\ref{ccc2}) the trace ${\rm Tr} (...)$ is over spin indices.
The Fermi functions $f_a$ and $f_b$ are arbitrary.\cite{but92}
Choosing $f_a=f_b=f_3$, we obtain
\begin{multline}
S_{23}(0) = -\frac{2e^2}{h} \sum_\sigma \int d\epsilon \,
\{ \tilde{T}_{1 2}^\sigma \tilde{T}_{1 3}^\sigma 
[f_1 -f_3 ]^2\\
+ \tilde{R}_{2 2}^\sigma  \tilde{T}_{3 2}^\sigma 
[f_2 -f_3 ]^2
+2\tilde{T}_{1 2}^\sigma  \tilde{T}_{2 3}^\sigma 
[f_1 -f_3 ] [f_2 -f_3 ] \}
\,,
\label{ccc3}
\end{multline}
where $R_{2 2}^\sigma$ is the reflection probability 
(in general $R_{\alpha\alpha}=1-\sum_\beta {T}_{\alpha\beta}$).
Notice that generally one cannot write the multilead
current--current correlations only
in terms of transmission probabilities as in Eq.~(\ref{ccc3}).
This was firstly pointed out by B\"uttiker,\cite{but90}
suggesting the appearance of exchange effects in noise measurements.
Here, since we are dealing with a (renormalized) Breit-Wigner resonance,
exchange corrections due to phase differences do not play any role.

For completeness, we give now the formula for the {\em shot noise}, i.e.,
the current-current correlations measured at the same lead
(e.g., lead 1). Following the way of reasoning that led
to Eq.~(\ref{ccc3}) we obtain
\begin{multline}
S_{11}(0) = \frac{2e^2}{h} \sum_\sigma \int d\epsilon \,
\{ \tilde{T}_{1 2}^\sigma \tilde{R}_{1 1}^\sigma 
[f_1 -f_2]^2\\
+ \tilde{T}_{1 3}^\sigma \tilde{R}_{1 1}^\sigma 
[f_1 -f_3 ]^2
+\tilde{T}_{1 2}^\sigma  \tilde{T}_{1 3}^\sigma 
[f_2 -f_3 ]^2 \}
\,.
\label{ccc4}
\end{multline}

\section{Results}
\label{results}
In the following, we present results obtained by self-consistently
solving Eqs.~(\ref{eomb}) and~(\ref{constraint}) for each bias voltage.
The rest of parameters is changed in the next subsections.
Throughout this work, we have checked that current conservation
($I_1+I_2+I_3=0$) is fulfilled.\cite{screening}

Tunneling effects are incorporated
at all orders since at equilibrium the Kondo temperature is found to be
\begin{equation}
\label{tkondocero}
k_B T_K^0=\tilde{\Gamma}=
D\exp{(-\pi|\varepsilon_0|/2\Gamma)}\,,
\end{equation}
which is clearly a nonperturbative result. In Eq.~(\ref{tkondocero})
$\Gamma=\sum_{\alpha=1}^{3}\Gamma_\alpha$ is the total hybridization broadening.
The reference energy will be always set at $E_F=0$ and the energy
cutoff is $D=100\Gamma$. The bare level is $\varepsilon_0=-6\Gamma$, deep
below $E_F$ to ensure a pure Kondo regime.

\subsection{Dephasing}
Before turning to the determination of current cross correlators,
we briefly discuss with an application the capabilities of
{\em three}-terminal setups to illustrate some difficult aspects
of the physics of the {\em two}-terminal Kondo effect.
As mentioned in the Introduction,
we investigate the action of a fictitious voltage probe\cite{but86b}
(say, lead 3) in order to {\em simulate} decoherence effects on the formation
of the Kondo resonance between leads 1 and 2.\cite{sil03}
These contacts play the role of source and drain, respectively.
The voltage probe model\cite{but86b}
describes decoherence since an electron that is absorbed
into the probe looses its coherence. The exiting
electron is replaced by an electron (with an unrelated phase)
injected by the probe.

At low temperatures the principal source of dephasing
is due to quasi-elastic scattering.\cite{but88} We consider then a voltage
probe that preserves energy.\cite{jon96} The current through the voltage
probe is zero at every energy $\epsilon$.
Thus, from Eq.~(\ref{cur2}) the distribution function at the probe reads
\begin{equation}
f_3(\epsilon)=\frac{T_{13}(\epsilon)f_1(\epsilon)
+T_{23}(\epsilon)f_2(\epsilon)}{T_{13}(\epsilon)
+T_{23}(\epsilon)} \,.
\label{f3}
\end{equation}
We have to insert this result into Eqs.~(\ref{eomb}) and~(\ref{constraint})
and solve self-consistently for
the hybridization couplings $\tilde{\Gamma}$
and the resonance level $\tilde{\varepsilon}_0$
in the presence of quasi-elastic scattering for each value of the applied bias voltage.
Then we compute numerically the differential conductance $G=dI/dV_{\rm sd}$,
where $I=I_1=-I_2$ and $V_{\rm sd}=V_1-V_2$.
\begin{figure}
\centerline{
\epsfig{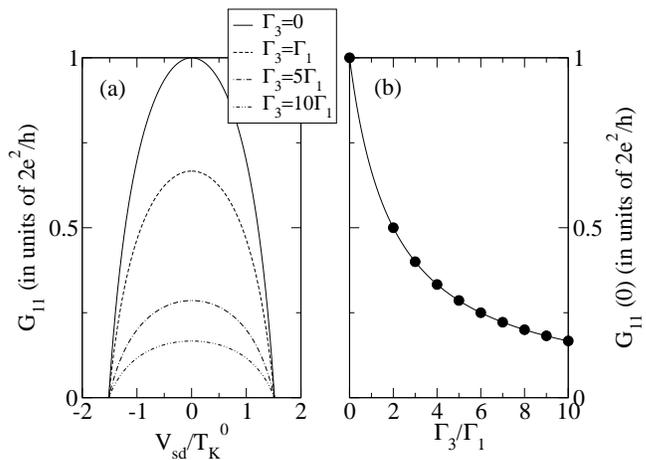}
}
\caption{(a) Differential conductance $G_{11}$ versus bias voltage
$V_{1}$ as a function of the bare coupling $\Gamma_3$ to the
voltage probe (reservoir 3) for $\Gamma_1=\Gamma_2$ and
$\varepsilon_0=-6\Gamma$.
(b) Linear conductance $G_{11}(0)$ showing the reduction of the peak
in (a) versus the coupling to the voltage probe. The dots
are numerical results where the line corresponds to an
analytical formula (see text).
}
\label{fig01}
\end{figure}

Figure~\ref{fig01}(a) shows $G$ for different values of the
coupling to the probe (we set $\Gamma_2=\Gamma_1$) .
For $\Gamma_3=0$ we obtain the well known zero-bias anomaly,
which arises from the formation
of the Kondo resonance at $V_{\rm sd}=0$.
As $\Gamma_3$ increases we observe a quenching of the Kondo peak.
The degree of the conductance suppression depends
on the coupling to the probe.
At each bias, $\mu_3$ (which has to be self-consistently calculated)
adjusts itself to fulfill the condition of zero net current
at each energy $\epsilon$. 
Hence, $\Gamma_3$ is a phenomenological parameter that includes
dephasing processes present in the quantum dot.
To see this, we can write down the current through, say, lead 1,
using Eqs.~(\ref{cur2}) and~(\ref{f3}):
\begin{equation}
I= \frac{e}{\hbar} \frac{4\tilde{\Gamma}_1\tilde{\Gamma}_2}
{\tilde{\Gamma}_1+\tilde{\Gamma}_2}
\int d\epsilon \, A_0 (\epsilon)
[f_1(\epsilon)-f_2(\epsilon)] \,,
\label{cur_1}
\end{equation}
where $A_0(\varepsilon)=-{\rm Im} \, G_0^r(\varepsilon)/\pi$
is the LDOS in the dot. Equation~(\ref{cur_1}) has the form of a
formula for a two-terminal current\cite{mei92}
with $G_0^r(\varepsilon)=[\varepsilon-\tilde{\varepsilon}_0
+i(\tilde{\Gamma}_1+\tilde{\Gamma}_2+\tilde{\Gamma}_3)]^{-1}$.
It is straightforward to show that a nonzero
${\Gamma}_3$ leads to deviations of Eq.~(\ref{cur_1})
from the unitary limit.

In Fig.~\ref{fig01}(b) we plot the linear conductance
$\mathcal{G}=G(V_{\rm sd}=0)$ as a function of $\Gamma_3/\Gamma_1$
from the results found numerically.
At zero bias we can find from Eq.~(\ref{cur_1})
an analytical expression for the reduction
of the peak:
\begin{equation}
\mathcal{G}=\frac{2e^2}{h} \frac{2}{2+\Gamma_3/\Gamma_1} \,.
\label{linearG}
\end{equation}
It is shown in Fig.~\ref{fig01}(b) (full line). In the limit of $\Gamma_3/\Gamma_1\ll 1$ a similar expression for the reduction of the peak was found by Kaminski {\it et al.},~\cite{kam00} the source of decoherence being an ac voltage applied to the dot level. 
 
\begin{figure}
\centerline{
\epsfig{file=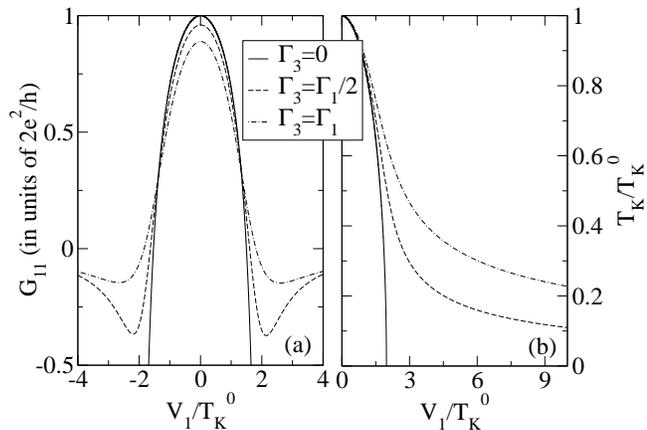,angle=270,width=0.47\textwidth,clip}
}
\caption{(a) Differential conductance $G_{11}$ versus bias voltage
$V_{1}$ for $\varepsilon_0=-6\Gamma$ ($T_K^0=8\times 10^{-3}\Gamma$).
(b) Dependence of the Kondo temperature on $V_{1}$.
}
\label{fig1}
\end{figure}

\subsection{Multiterminal conductance}\label{sec_mc}
From now on, we consider lead 3 as a real electrode with tunable voltage
$V_3$. We set $V_3=V_2=0$ and vary the tunneling coupling $\Gamma_3$.
The self-consistent results of Eqs.~(\ref{eomb}) and~(\ref{constraint})
are inserted in Eqs.~(\ref{cur2})
to calculate the differential conductance through lead 1:
$G_{11}=dI_1/dV_{1}$. Figure~\ref{fig1}(a) shows $G_{11}$
as a function of $V_1$. At $\Gamma_3=0$ the conductance at $V_1=0$
achieves the unitary limit as in the two-terminal case.
With increasing the coupling to third lead,
$G_{11}(0)$ decreases. For equal tunnel couplings
($\Gamma_1=\Gamma_2=\Gamma_3=\Gamma/3$),
$G_{11}(0)$ does not reach
1 (in units of $2 e^2/h$) but instead $G_{11}(0)=8/9$, in agreement
with Ref.~\onlinecite{cho03}.
This is an immediate consequence of having three leads
with identical couplings.
Interestingly, the Kondo temperature of Fig.~\ref{fig1}(b) does {\em not}
vanish abruptly for $V_1=2 T_K^0$, as known in the two-terminal case
(see the case $\Gamma_3=0$).
This is an important result as it
implies that Kondo correlations survive at large voltages.
The effect is reminiscent of the situation found by Aguado and Langreth\cite{agu00}
in tunnel-coupled double quantum dots, though the physical origin
is clearly distinct.

\subsection{Sign of the current cross-correlations.
Comparison with a noninteracting quantum dot}\label{sec_cc}
\begin{figure}
\centerline{
\epsfig{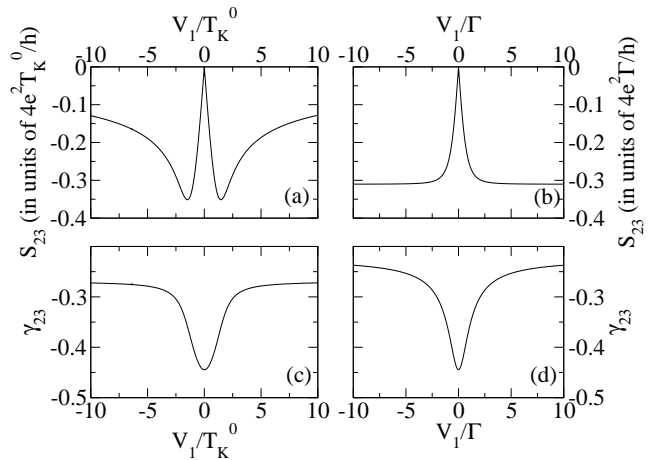}
}
\caption{(a) Current-current cross correlation measured in leads
2 and~3, $S_{23}(0)$, as a function of the bias voltage in the injecting
lead, $V_1$. Kondo correlations involve an increase of $S_{23}(0)$
for voltages larger than $2T_K$.
(b) Same as (a) for a noninteracting quantum dot with a resonant level
exactly at $E_F$. (c) and (d) correspond to the
Fano factor $\gamma_{2 3}$ as a function of
voltage for the interacting and noninteracting case respectively.
}
\label{fig2}
\end{figure}

We now focus on the current-current correlations of the current
for $V_3=V_2=0$ and equal couplings $\Gamma_1=\Gamma_2=\Gamma_3=\Gamma/3$.
Later, we shall allow for nonzero voltage differences between leads 2 and 3.
In Fig.~\ref{fig2}(a), we show the cross correlator $S_{23}(0)$ obtained
from Eq.~(\ref{ccc3}). As expected, $S_{23}$ is zero
for $V_1=0$ and negative elsewhere.
This reflects the fermionic nature of the quasiparticles.
For comparison, we plot in Fig.~\ref{fig2}(b) the corresponding $S_{23}$
for a noninteracting resonant double-barrier structure
with the level at $E_F$
(of course, for $\varepsilon_0=-6\Gamma$ the spectrum
$S_{23}$ is always very small as the transmission is).
In this case, the physics is governed by the bare coupling
$\Gamma$.\cite{che91} On the contrary,
in the Kondo problem the dominating energy scale is $T_K$.
Qualitatively, Fig.~\ref{fig2}(a) and~\ref{fig2}(b)
look the same until $V_{1}\sim 2T_K$.
The cross correlator in the Kondo case increases with voltage while
in the noninteracting case $S_{23}$ saturates at large voltages.
It is easy to show that the saturation value is given by $-8\pi/81\simeq -0.31$
(in units of $4e^2\Gamma/h$). The reason for the increase of
$S_{23}(0)$ in Fig.~\ref{fig2}(a) is that $T_K$ is voltage dependent
unlike the bare $\Gamma$, even in the wide-band limit.
In particular, the current--voltage characteristics shows a region
of negative differential conductance in the Kondo case [see Fig.~\ref{fig1}(a)]
whereas it reaches a constant value at large voltages
for an noninteracting quantum dot.

To avoid effects due to moderate biases, in what follows we shall concentrate
on a normalized $S_{23}$. We define the Fano factor of $S_{23}$ as
\begin{equation}
\gamma_{2 3}=\frac{S_{2 3}}{2 e \sqrt{|I_2||I_3|}} \,.
\end{equation}
If the scattering region were a simple barrier of transmission
$T$, $\gamma_{2 3}$ would be simply $-1$. This number changes
when the system under consideration is a quantum dot.
In Figs.~\ref{fig2}(a) and~(b), we plot $S_{2 3}$
for the Kondo and the noninteracting case, respectively.
Their corresponding Fano factors are shown in
Figs.~\ref{fig2}(c) and~(d).
We see that $\gamma_{2 3}$ has a minimum
at $V_1=0$. Analytically, we find $\gamma_{2 3}(0)=-4/9\simeq -0.44$, which
is in excellent agreement with the numerical result.
Likewise, we can assess the limit of $\gamma_{2 3}$ at very high voltages
($V_1\gg T_K^0$). We get $\gamma_{2 3}=-2/9\simeq -0.22$.
As observed, both curves tend to this value, though for a noninteracting
quantum dot it is more quickly due to the independence of $\Gamma$ on the
bias voltage.
\begin{figure}
\centerline{
\epsfig{file=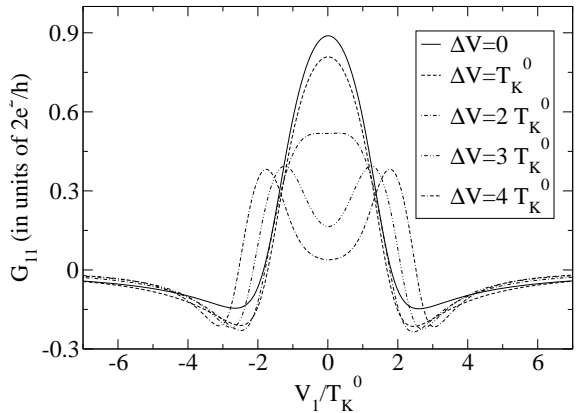,angle=270,width=0.47\textwidth,clip}
}
\caption{Differential conductance $G_{11}$ versus bias voltage
$V_{1}$ for different values of the voltage difference
$\Delta V\equiv |V_2-V_3|\gtrsim 2T_K^0$.
}
\label{fig4}
\end{figure}

\subsection{Effect of the nonequilibrium splitting on the current-current
correlations}
Now we turn to an exciting case. Consider the following bias configuration:
$V_2=-V_3\neq 0$ and determine the differential conductance $G_{1 1}$
as a function of $V_1$. The case $V_2=-V_3=0$ has been treated before.
However, due to the fact that the boson field never vanishes, we can
study the situation $\Delta V\equiv |V_2-V_3|\gtrsim 2T_K^0$.
As remarked in the Introduction,
it has been argued\cite{sun01,leb01}
and experimentally observed\cite{fra02} that in a three-lead geometry
the splitting of the Kondo resonance due to voltage is visible,
unlike the two-terminal case.
Moreover, in Refs.~[\onlinecite{leb01,cho03}] it has been noticed
that the conductance $G_{11}$ is not sensitive to the strength of the coupling to
the third lead, showing always a two-peak structure. 
Of course, only when the third lead is weakly coupled to the dot  $G_{11}$ 
is a measure of the LDOS. But since we are interested in the transport properties
of the system, our choice of equal coupling constants does not affect the results for
the conductances and the current-current correlations.

 In Fig.~\ref{fig4} we plot the behavior of the differential conductance
$G_{11}$. At $\Delta V=0$ we obtain the zero-bias
anomaly of Fig.~\ref{fig1}(a). As $\Delta V$ increases, $G_{1 1}$
is split at $V_2\sim T_K^0$.
Both splitting peaks lie at $V_1\sim V_3$
and $V_1\sim V_2$, i.e., when a pair of electrochemical potentials
are aligned. It is also at those points where the Kondo temperature is larger.
We emphasize that this effect has {\em no}
similitude in the electronic transport
through a noninteracting quantum dot.
Still, a mean-field theory of the Kondo effect as presented here
is able to capture this physics.
At the same time that the splitting in $G_{11}$ develops,
the height of the peaks decreases, suppressing
the zero-bias anomaly, although not so strongly as in the experiment\cite{fra02} 
due to the absence of inelastic scattering in this case.
\begin{figure}
\centerline{
\epsfig{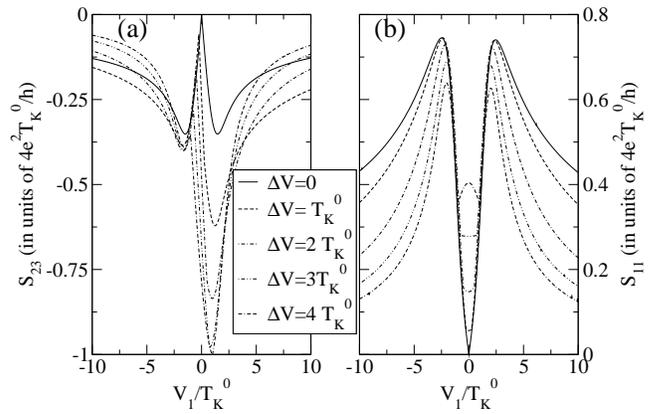}
}
\caption{(a) Cross correlations of the current measured
between leads 2 and~3 for the case treated
in Fig.~\ref{fig4}. (b) Same as (a) for the shot noise in lead 1.
}
\label{fig45}
\end{figure}

We now use Eq.~(\ref{ccc3}) to calculate the cross correlations
between leads 2 and 3. The results are presented in Fig.~\ref{fig45}(a).
The dependence of $S_{2 3}$ on voltage is rather asymmetric, hindering
the observation of a clear indication due to the voltage induced splitting.
The asymmetry is caused by the third term of the right-hand side of
Eq.~(\ref{ccc3}), which is not symmetric under the operation $V_1\to -V_1$
when $\Delta V > 0$. That is the reason why we next consider the shot noise
in lead 1 $S_{11}$, which is
an even function of the applied $V_1$.

In Fig.~\ref{fig45}(b) we plot the results of Eq.~(\ref{ccc4}).
We observe that $S_{11}$ at $V_1=0$ is nonvanishing with
increasing $\Delta V$, causing a {\em divergence} of the Fano factor.
This is not related to the Kondo physics but with the fact
that the lead 1 at $V_1=0$ acts as a voltage probe with
{\em zero} impedance since the net current flowing through it is zero.
Including the fluctuations of the potentials would probably
cancel out the divergence. A consequence of Kondo physics
is that the minimum at $V_1=0$ turns into a maximum. This occurs
when the splitting in $G_{11}$ is sharply formed
[see Fig.\ref{fig4}].

\subsection{Spin dependent transport and current cross correlations}
So far we have assumed spin-independent transport.
Let us go back to the bias configuration of Secs.~\ref{sec_mc} and~\ref{sec_cc}
($V_2=V_3=0$) and focus on the spin-dependent transport properties.
It is customary in the theoretical studies of spintronic transport
to take into account the influence of external magnetic fields and
ferromagnetic electrodes, among other parameters.\cite{wol01}
Firstly, we shall change the external Zeeman field
and then enable the presence of spin-polarized tunneling. 

\subsubsection{Magnetic field}
We assume that the leads are paramagnetic and that
the magnetic field is applied only to the dot, resulting in
a Zeeman gap of the bare resonant level:
$\Delta_Z=\varepsilon_{0\uparrow}-\varepsilon_{0\downarrow}$.
It is well known that, as a consequence, the Kondo resonance is
split when $\Delta_Z\sim T_K^0$.\cite{mei93}

\begin{figure}
\centerline{
\epsfig{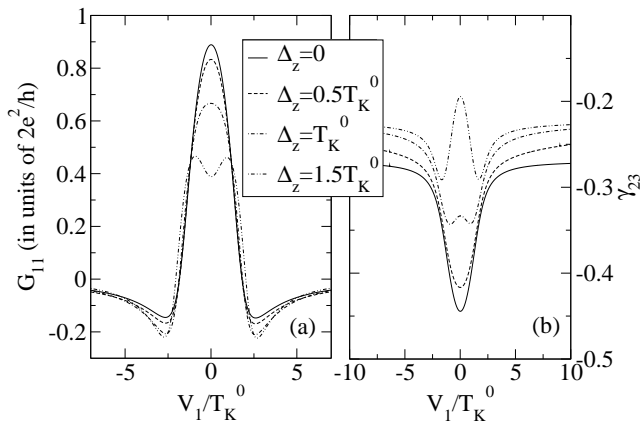}
}
\caption{(a) Differential conductance $G_{1 1}$ versus $V_{11}$
as a function of the Zeeman term $\Delta_Z$ for $V_2=V_3=0$.
(b) Same as (a) for the Fano factor of the cross correlator, $\gamma_{2 3}$.}
\label{fig5}
\end{figure}

Figure~\ref{fig5}(a) shows the differential conductance $G_{1 1}$
for different values of the Zeeman field. The conductance
is split and quenched with increasing $\Delta_Z$, as expected.
In Fig.~\ref{fig5}(b), we depict the Fano factor of the
cross correlator $\gamma_{2 3}$. It exhibits a very interesting
feature. Due to the splitting of the Kondo peak, the minimum of
the cross correlator at $V_1=0$ becomes a local maximum, resulting
from the suppression of the Kondo effect. However, this change
occurs {\em before} the splitting of the conductance $G_{1 1}$.
Therefore, measuring the shot noise provides {\em new}
information in this case. The presence of the splitting 
would be detected in an experiment more precisely by means
of the shot noise. The underlying reason is that the form of
Eq.~(\ref{ccc3}) differs from that of the current
which is basically proportional to $\tilde{T}_{12}$ alone,
see Eq.~(\ref{cur2}).
As a result, the width of the $G_{1 1}$ resonance is a bit larger
than the $\gamma_{2 3}$ antiresonance and the former is then
more robust than the latter against the application of magnetic fields.

\subsubsection{Ferromagnetic leads}
There has recently been considerable debate about the influence of
ferromagnetic leads in the Kondo physics of a quantum dot.\cite{serg,zhang,mar,choi03}
In the preceding subsection, it was clear that an external magnetic
field alters the real part of the quantum-dot self-energy, breaking
the spin degeneracy. In the case of spin polarized tunneling, the
situation is more subtle.\cite{choi03}
When the magnetic moments of the contacts are aligned along the same direction,
the density of states
of the localized electron undergoes a splitting if particle-hole
symmetry is broken.\cite{asy}
Recent transport experiments with C$_{60}$ molecules
and carbon nanotubes
have addressed this regime.\cite{pas04,nyg04}
However, in our case the dot is in the strong coupling limit and the
Kondo effect is pure in the sense that no charge fluctuations
are allowed. Thus, no splitting is expected in the differential conductance.
\begin{figure}
\centerline{
\epsfig{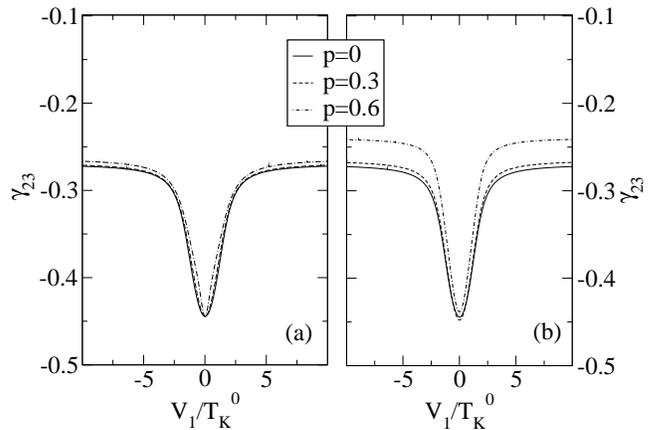}
}
\caption{Fano factor of the cross correlator, $\gamma_{2 3}$ vs $V_{1}/T_K^0$ for different 
lead magnetization  when $V_2=V_3=0$.  (a) Parallel alignment between the magnetizations of the leads with spin polarizations: $p_1=p_2=p_3=p$ . (b) Antiparallel case with  $p_1=-p_2=-p_3=p$.}
\label{fig6}
\end{figure}

In Fig.~\ref{fig6}(a), we show  the cross correlator $\gamma_{23}$ for different values of the lead magnetization in the parallel case. This means that $p_1=p_2=p_3=p$,
where $p_\alpha$ is the spin polarization of lead $\alpha$.
Ferromagnetism in the leads arises through spin-dependent
densities of states $\nu_{\alpha\sigma}(\epsilon)=\sum_k\delta
(\epsilon-\varepsilon_{k\alpha\sigma})$.
Hence, the linewidths become spin dependent:
$\Gamma_{\alpha\sigma}=(1 \pm p_\alpha)\Gamma_\alpha$, where +(-)
corresponds to up (down) spins.
We prefer to restrict $p_\alpha$ to small values as
strong magnetizations would require a proper treatment
of the reduction of the bandwidth $D$.
We observe that $\gamma_{23}$  is rather insensitive to changes in $p$ in the same fashion as
$G_{11}$ is in the Fermi-liquid fixed point\cite{lop03a}. Only at moderate polarizations ($p=0.6$) we see that the dip in $\gamma_{23}$ gets narrower because the Kondo temperature decreases as $p$ increases\cite{mar,choi03}. In addition, $\gamma_{23}$ is always negative in contrast to the results obtained in the Coulomb blockade regime where $\gamma_{23}$ can take positive values~\cite{cot03}. When the spin-flip scattering rate is smaller than the tunneling rate, $\gamma_{23}$ can be positive. However in the Kondo regime this condition is never met since the rate of spin flip scattering  $\sim 1/T_K$ is always much longer than the tunneling rate $\sim 1/\Gamma$. Figure~\ref{fig6}(b) is devoted to the antiparallel case: $p_1=-p_2=-p_3=p$. Accordingly, $\gamma_{23}$ is lifted with increasing
lead polarization since the conductance peak decreases
with increasing $p$ (roughly, with a factor $1-p^2$)~\cite{lop03a}. 

\section{Conclusion}\label{conclusion}
In summary, we have investigated the Kondo temperature,
the differential conductance and cross correlations of the current
when three leads are coupled to an artificial Kondo impurity
in the Fermi-liquid fixed point of the infinite-$U$
Anderson Hamiltonian
($T\ll T_K$). We have performed a systematic study
of the properties of the cross correlators when dc bias, 
Zeeman splittings, and ferromagnetic leads influence the
nonequilibrium transport through the quantum dot.
Our most relevant result is the behavior of the shot noise
when there arises a voltage induced splitting in the quantum dot.

In addition, we have studied the current of a two-terminal
quantum dot attached to a voltage probe. We have shown
that increasing the coupling with the probe induces
a quenching of the Kondo peak.
Despite the simplicity of this approach, it gives rise to
results that are in agreement with more sophisticated models,\cite{kam00,lop01}
though the precise processes responsible for the decoherence
need still to be derived from a microscopic model.

We have not exhausted all the possibilities that the model offers
and more complicated geometries with appealing results can be envisaged.
One could address the situation with two injecting and two receiving
leads, which could give rise to Hanbury Brown-Twiss-like effects.\cite{hbt}
We expect that phase related exchange terms will arise
especially at higher temperatures ($T>T_K$),
when the singlet state between the localized spin and
the conduction electrons is not yet well formed.
We believe that in the presence of spin-polarized couplings due to
ferromagnetic leads, bunching effects will be enhanced.\cite{nextwork}

Improvements of the model should go in the direction
of including fluctuations of the boson field and of the
renormalized level. However, we do not expect large deviations
from the results reported here when $T\ll T_K$.
These fluctuations will evidently become important as
temperature approaches $T_K$.
Experimentally, our predictions can be tested with
present technology such as GaAs quantum dots\cite{fra02}
or carbon-nanotube nanostructures.\cite{par02}

\section*{Acknowledgements}
We gratefully acknowledge R. Aguado, M. B\"uttiker, S. Pilgram and
P. Samuelsson for helpful comments.
This work was supported by the EU
RTN under Contract No. HPRN-CT-2000-00144, Nanoscale Dynamics,
and by the Spanish MECD.


\begin{thebibliography}{90}
\bibitem{screening}
Since we take the limit $U\to\infty$,
we can safely neglect screening effects
(the charge is fixed). However, in a realistic situation one should
take into account a screening potential,
which may become important in the nonlinear regime.
\bibitem{asy}
This is associated with the asymmetric Anderson model.
\end{thebibliography}
\end{document}